\documentstyle[12pt]{article}
 
\begin{document}
 \begin{titlepage}
 \begin{center}
 {\Huge Fusion rules in N=1}
\vspace*{1cm}

{\Huge superconformal minimal models}
 
 \vspace*{1cm}
 
 {\large Pablo Minces$^{\dag, *}$, M.A. Namazie$^{\ddag}$ and  Carmen
N\'u\~nez$^{\dag}$}
 \vspace*{0.5cm}
 
 $^{\dag} $Instituto de Astronom\'{\i}a y F\'{\i}sica del Espacio
(CONICET)\\ 
C.C.67 - Suc.28, 1428 Buenos Aires,
 Argentina \\
$^*$ Universidade de S\~ao Paulo, Instituto de F\'{\i}sica\\
Caixa Postal 66.318 - CEP 05315-970 - S\~ao Paulo - Brasil\\
$^{\ddag}$ namazie@pacific.net.sg
 \vspace*{1cm}
 
 \end{center}
 
 \begin{abstract}
 The generalization to N=1 superconformal minimal
 models of the relation between the modular transformation
 matrix and the fusion rules in rational conformal field theories, the
 Verlinde theorem, 
 is shown to provide complete information about the fusion rules,
 including their fermionic parity.  The results for the superconformal
Tricritical Ising and Ashkin-Teller models agree with the known rational
 conformal
 formulation.
 The Coulomb gas description of correlation functions in the Ramond
sector
 of N=1 minimal models is also discussed and a previous formulation is
 completed.
 \end{abstract}
 
 \end{titlepage}

 \section{Introduction.}
 
 One of the most interesting results in 2D conformal field theory is the
 Verlinde theorem \cite{verlinde}\cite{verlinde2}\cite{moore}. It gives
the
 number of conformal blocks of a RCFT on a punctured Riemann surface in 
terms of elements of the modular matrix. The
 result arises as a consequence of the well established, but nevertheless
 still surprising, fact that the modular matrix S implementing the modular
 transformation $\tau \rightarrow -1/\tau $ on the space of genus one
 conformal blocks, diagonalizes the fusion rules. The proof of the
theorem
 relies on the technical assumption that both left and right extended chiral
 algebras consist only of generators with integral conformal weight, 
the so called rational conformal field theories (RCFT's), and thus
 evidently excludes the superconformal case.
 
 A generalized Verlinde formula which describes the fusion rules in all sectors
of $%
 N=1$ superconformal theories was obtained in \cite{aann}: by
considering
 some examples of correlation functions of primary fields and employing
 certain bases of conformal blocks, it was argued that the Verlinde
 conjecture extends to the $N=1$ superconformal unitary series. In this
 letter we show that the generalized Verlinde formula contains information
 about the fermionic parity of the fusion rules. 

In Section {\bf {2.}} we complete the formulation of the Coulomb gas
description \cite{dotsenko}
of
 correlation functions in the $R$ sector performed in \cite{jnnns} and
\cite
 {aann}, by considering the parity of the R fields. In Section {\bf {3.}}
we
 discuss the generalized Verlinde formula and compute the fusion rules of
the
 Tricritial Ising model (TIM) and the critical Ashkin-Teller model (AT). These
models
 are known to have both RCFT as well as superconformal descriptions, so they
 can be regarded as a check of our results. Since the superconformal 
Coulomb gas
method
 remains at the level of a prescription, the correlation functions have to
 pass several checks such as null state decoupling and correct
behavior
 under degeneration of the torus to the plane as well as in the
factorization
 limit. Consistency with the conformal fusion rules can therefore be
 considered another successful check on the superconformal blocks and
on
the
 extension of the Verlinde formula.
 
 \section{Two-Point Conformal Blocks.}
 
 Contour integral representations of the conformal blocks can be computed
 with the Dotsenko-Fateev Coulomb gas technique \cite{dotsenko} by
 introducing Feigin-Fuks screening operators to make correlation functions
 background charge neutral. Correlation functions of $N=1$ superconformal
 primary fields on the torus were considered in \cite{jnnns}\cite{aann}
where
 one contour examples, corresponding to Ramond primary fields, and double
 contour integrals, necessary when considering NS primary fields, were
 studied. In this section we rely heavily on notation, results and
 the discussion contained in these references.
 
 The N=1 superconformal minimal models are characterized by the
following
 discrete series of central charges $c$ and allowed conformal weights of
the
 primary fields $\Delta _{r,s}$ \cite{friedan}.
 
 \begin{eqnarray}
 c &=&\frac 32-\frac{12}{p(p+2)}\qquad (p=3,4,...)  \nonumber \\
 \Delta _{r,s} &=&\frac{[(p+2)r-sp]^2-4}{8p(p+2)}+\frac
1{32}[1-(-1)^{r-s}]
 \label{e1}
 \end{eqnarray}
 where $1\leq s\leq r\leq p-1$ with $r-s$ even in the NS sector and $1\leq
 s\leq r-1$ for $1\leq r\leq \left[ \frac{p-1}2\right] $ or $1\leq s\leq
r+1$
 for $\left[ \frac{p+1}2\right] \leq r\leq p-1$ with $r-s$ odd in the R
 sector.
 
 In order to compute correlation functions, the fields are represented by
 vertex operators of the form \cite{mussardo}
 
 \begin{equation}
 N_\alpha ^B(z)=e^{i\alpha \phi (z)}\qquad N_\alpha ^F(z)=\psi
(z)e^{i\alpha
 \phi (z)}\qquad R_\alpha ^{\pm }(z)=\sigma ^{\pm }(z)e^{i\alpha \phi (z)}
 \label{e27}
 \end{equation}
 where $\phi $ and $\psi $ are a free boson and a Majorana fermion,
$N_\alpha
 ^B$ and $N_\alpha ^F$ are the bosonic and fermionic components of a $NS$
 field, $\sigma ^{\pm }$ are the two spin fields with conformal weight
$\frac
 1{16}$ that are to be identified respectively with the spin field and the
 disorder field of the Ising model, and $R_\alpha ^{\pm }$ are the two $R$
 fields whose conformal weights correspond to the charge $\alpha $ (only
the
 ground states of conformal weight $\frac c{24}$ are not degenerate). They
 obey the following OPE 
 \begin{eqnarray}
 \psi (z)\sigma ^{\pm }(w) &\sim &\frac 1{(z-w)^{\frac 12}}\sigma ^{\mp
 }(w)+...  \nonumber  \label{3} \\
 \sigma ^{\pm }(z)\sigma ^{\pm }(w) &\sim &\frac 1{(z-w)^{\frac
 18}}+(z-w)^{\frac 38}\psi (w)+...  \nonumber \\
 \sigma ^{\pm }(z)\sigma ^{\mp }(w) &\sim &(z-w)^{\frac 38}\psi (w)+...
 \label{e27}
 \end{eqnarray}
 
 Let us compute the conformal blocks corresponding to the correlator $%
 \left\langle \phi _{1,2}\phi _{1,2}\right\rangle $ using the Coulomb gas
 formalism and completing the results in \cite{jnnns} and \cite{aann} by
 taking into account the fermionic parities of the fields, i.e.,
considering
 the three possible two-point functions $\left\langle \phi _{1,2}^{+}\phi
 _{1,2}^{+}\right\rangle $, $\left\langle \phi _{1,2}^{-}\phi
 _{1,2}^{-}\right\rangle $ and $\left\langle \phi _{1,2}^{+}\phi
 _{1,2}^{-}\right\rangle $. The conformal blocks are
 
 \begin{equation}
 G_i^{(sign1,sign2);\nu }=\oint\limits_{C_i}dz~\left\langle \sigma
 ^{sign1}(z_1)\sigma ^{sign2}(z_2)\psi (z)\right\rangle _\nu \left\langle
 e^{-\frac i2\alpha _{-}\phi (z_1)}e^{-\frac i2\alpha _{-}\phi
 (z_2)}e^{i\alpha _{-}\phi (z)}\right\rangle _\nu  \label{e28}
 \end{equation}
 where $\nu =1,2,3,4$ label the spin structure and
 
 \begin{equation}
 \alpha _{+}=\sqrt{\frac{p+2}{2p}}\qquad \alpha _{-}=-\sqrt{\frac
p{2(p+2)}}
 \label{e28 2}
 \end{equation}
 
 The closed contour basis $\left\{ C_1,C_2\right\} $ has been specified
in 
 \cite{jnnns}\cite{aann}, and $sign1,sign2\in \left\{ +,-\right\} $.
 
 The correlation function of $2n$ spin fields and $2m$ disorder fields of
the
 Ising model on the torus is given by \cite{di francesco}
 
 \[
 \left\langle \sigma ^{+}(w_1,\bar w_1)...\sigma ^{+}(w_{2n},\bar
 w_{2n})\sigma ^{-}(u_1,\bar u_1)...\sigma ^{-}(u_{2m},\bar
 u_{2m})\right\rangle _\nu ^2= 
 \]
 
 \begin{eqnarray}
 \ &=&\sum\limits_{\stackrel{\varepsilon _i=\pm 1}{\stackrel{\varepsilon
 _k^{\prime }=\pm 1}{\sum \varepsilon _i+\sum \varepsilon _k^{\prime
}=0}}%
 }\left| \frac{\Theta _\nu \left( \frac{\sum \varepsilon _iw_i+\sum
 \varepsilon _k^{\prime }u_k}2\right) }{\Theta _\nu (0)}\right|
 ^2\prod\limits_k\varepsilon _k^{\prime
}\prod\limits_{\stackrel{i<j}{k<l}%
 }\left| \frac{\Theta _1(w_i-w_j)}{\Theta _1^{\prime }(0)}\right|
^{\frac{%
 \varepsilon _i\varepsilon _j}2}  \nonumber \\
 &&\ \times \left| \frac{\Theta _1(u_k-u_l)}{\Theta _1^{\prime
}(0)}\right| ^{%
 \frac{\varepsilon _k^{\prime }\varepsilon _l^{\prime
}}2}\prod\limits_{i,k}%
 \left| \frac{\Theta _1(w_i-u_k)}{\Theta _1^{\prime }(0)}\right| ^{\frac{%
 \varepsilon _i\varepsilon _k^{\prime }}2}  \label{e29}
 \end{eqnarray}
 where $\Theta _\nu $ are the usual Jacobi $\Theta $-functions.
 
 We take the square root on both sides of this equality and keep the
 holomorphic part only (obtaining in this way the so called `holomorphic
 square root'). We then specialize to $n=2,$ $m=0;$ $n=0,$ $m=2$ and
$n=m=1$
 and take the limits $w_3\rightarrow w_4$, $u_3\rightarrow u_4$ and $%
 w_2\rightarrow u_3$ respectively. In all cases (because of (\ref{e27}))
the
 residue of the $\frac 38$ order pole is kept. We thus obtain
 
 \begin{eqnarray}
 \left\langle \sigma ^{+}(z_1)\sigma ^{+}(z_2)\psi (z)\right\rangle _\nu
 &=&\left\langle \sigma ^{-}(z_1)\sigma ^{-}(z_2)\psi (z)\right\rangle
_\nu 
 \nonumber \\
 &=& \left[ \frac{\Theta _1(z_1-z_2)}{\Theta _1^{\prime }(0)}\right]
^{\frac
 38}\left[ \frac{\Theta _1(z_1-z)}{\Theta _1^{\prime }(0)}\right] ^{-\frac
12}
 \nonumber \\
 &&\times \left[ \frac{\Theta _1(z_2-z)}{\Theta _1^{\prime }(0)}\right]
 ^{-\frac 12}\left[ \frac{\Theta _\nu \left( \frac 12(z_1-z_2)\right) }{%
 \Theta _\nu (0)}\right] ^{-\frac 12}  \nonumber  \label{mas mas} \\
 &&\times \frac{\Theta _\nu \left( \frac 12(z_1+z_2-2z)\right) }{\Theta
_\nu
 (0)}  \label{e30}
 \end{eqnarray}
 
 \begin{eqnarray}
 \left\langle \sigma ^{+}(z_1)\sigma ^{-}(z_2)\psi (z)\right\rangle _\nu
 &=&\left[ \frac{\Theta _1(z_1-z_2)}{\Theta _1^{\prime }(0)}\right]
^{-\frac
 18}\left[ \frac{\Theta _1(z_1-z)}{\Theta _1^{\prime }(0)}\right] ^{-\frac
 12}\left[ \frac{\Theta _1(z_2-z)}{\Theta _1^{\prime }(0)}\right] ^{-\frac
12}
 \nonumber \\
 &&\times [\frac{\Theta _1^{\prime }(z_1-z)}{\Theta _1^{\prime
}(0)}\frac{%
 \Theta _1(z_2-z)}{\Theta _1^{\prime }(0)}\frac{\Theta _\nu \left( \frac
 12(z_1-z_2)\right) }{\Theta _\nu (0)}  \nonumber \\
 &&-\frac{\Theta _1(z_1-z)}{\Theta _1^{\prime }(0)}\frac{\Theta _1^{\prime
 }(z_2-z)}{\Theta _1^{\prime }(0)}\frac{\Theta _\nu \left( \frac
 12(z_1-z_2)\right) }{\Theta _\nu (0)}  \nonumber \\
 &&-2\frac{\Theta _1(z_1-z)}{\Theta _1^{\prime }(0)}\frac{\Theta
_1(z_2-z)}{%
 \Theta _1^{\prime }(0)}\frac{\Theta _\nu ^{\prime }\left( \frac
 12(z_1-z_2)\right) }{\Theta _\nu (0)}]^{\frac 12}  \label{e32}
 \end{eqnarray}
 
 By calculating the lattice contribution in (\ref{e28}) we get
 
 \begin{eqnarray}
 G_i^{(sign1,sign2);\nu }(r,s) &=&\oint\limits_{C_i}dz~\left\langle \sigma
 ^{sign1}(z_1)\sigma ^{sign2}(z_2)\psi (z)\right\rangle _\nu  \nonumber
 \label{bloques superconformes 2} \\
 &&\times \left[ \frac{\Theta _1(z_1-z_2)}{\Theta _1^{\prime }(0)}\right]
 ^{\frac 12\alpha _{-}^2}\left[ \frac{\Theta _1(z_1-z)\Theta _1(z_2-z)}{%
 \Theta _1^{\prime }(0)^2}\right] ^{-\alpha _{-}^2}  \nonumber \\
 &&\times \left[ \Gamma ^\nu (\lambda )-(-1)^{r\delta _{1\nu
}+r(s+1)\delta
 _{4\nu }}\Gamma ^\nu \left( \tilde \lambda \right) \right]  \label{e34}
 \end{eqnarray}
 where
 
 \begin{eqnarray}
 \Gamma ^\nu (\lambda ) &=&\frac{\Theta _\nu (0)^{\frac 12}}{\eta ^{\frac
32}}
 e^{i\pi \left( \frac{\lambda ^2}{2N}-\frac 1{12}\right) \delta _{4\nu
 }}\sum\limits_{n\in {\bf {Z}}}(-1)^{np(\delta _{1\nu }+\delta _{4\nu })} 
 \nonumber  \label{lattice} \\
 &&\ \times ~q^{\frac{(\lambda +nN)^2}{4N}}e^{i\pi \frac{\lambda
+nN}{\sqrt{N}%
 }\alpha _{-}(2z-z_1-z_2)}  \label{e35}
 \end{eqnarray}
 
 \begin{equation}
 \lambda =r(p+2)-sp\quad \quad \tilde \lambda =-r(p+2)-sp\quad \quad
N=2p(p+2)
 \label{e36}
 \end{equation}
 
 Notice that only the fermionic correlator (\ref{e30}) was considered in
\cite
 {jnnns}\cite{aann}. Clearly from eq. (\ref{e30}) $G_i^{(+,+);\nu
 }=G_i^{(-,-);\nu }$. Nevertheless, it is possible to see that both $%
 \left\langle \sigma ^{+}(z_1)\sigma ^{+}(z_2)\psi (z)\right\rangle _\nu $
 and $\left\langle \sigma ^{+}(z_1)\sigma ^{-}(z_2)\psi (z)\right\rangle
_\nu 
 $ have the same monodromy and modular properties as well as degeneration
and
 factorization limits. Indeed, in the degeneration limit,$q\rightarrow 0$,
 the four-point functions on the sphere are recovered, namely
 
 \begin{equation}
 \left\langle V_\alpha
(0)R_{1,2}^{sign1}(x_1)R_{1,2}^{sign2}(x_2)V_{2\alpha
 _0-\alpha }(\infty )\right\rangle  \label{e38}
 \end{equation}
 where the conjugate vertices $V_\alpha $ and $V_{2\alpha _0-\alpha }$ are
 either $N_\Delta ^B$, $N_\Delta ^F$ or $R_\Delta ^{\pm }$ and the
contours $%
 C_1$ and $C_2$ on the torus degenerate to Pochhammer contours on the
extended complex plane.
 
 The properties of the conformal blocks under monodromy transformations
are
 not altered. Recall that $\phi(a)$ and $\phi(b)$ are the Verlinde
operators
 which implement these transformations as the point $z_1$ (or $z_2$) is
 transported once around either an $a$ or $b$ cycle on the torus. These
 operators are not modified when the correlation functions contain spin
 fields of different parity.
 
 Another check on the conformal blocks is provided by the factorization
limit 
 $z_1\rightarrow z_2$, where the intermediate states are precisely those
 dictated by the fusion rules. The problematic feature appearing for even
$p$
 however, remains. In fact, the $N=1$ superconformal partition function
can
 always be obtained by factorizing the modular and monodromy invariant two
 point correlation function on the identity intermediate state. However in
 the $p=4$ case or AT model at criticality, the $C_1$ contour in the
 factorization limit reproduces all the terms of the corresponding
 superconformal partition function \cite{yang2}, except for the
contribution
 Tr$_{[R]}$(-1)$^{{F}}$ of the $\nu =1$ sector. This can be understood
since
 the intermediate state in the $\nu =1$ sector is a fermion, so we do not
 expect to obtain the partition function but the fermion expectation
value.
 Actually when $\nu =1$ the residue is zero, reflecting the fact that the
 fermion is a null state.
 
 Therefore taking into account the parities of the $R$ fields does not
modify
 the checks on the conformal blocks that have been performed previously in
\cite{jnnns}%
 \cite{aann}.
 
 Let us now consider the modular transformation matrix ${\bf {S}}$ on the
 basis of conformal blocks $G_{i=1}^{(sign1,sign2);\nu }(r,s)$. It is
 possible to see that $\left\langle \sigma ^{+}(z_1)\sigma ^{+}(z_2)\psi
 (z)\right\rangle _\nu $ and $\left\langle \sigma ^{+}(z_1)\sigma
 ^{-}(z_2)\psi (z)\right\rangle _\nu $ have the same $S$ transformation
 properties, so the form of the ${\bf {S}}$ matrix does not depend on the
 parity of the fields. Notice that the ${\bf S}_{\nu ^{\prime
}=1,2,3,4}^{-1}$
 matrices are just those given in \cite{aann}. We list them here (up to
phase
 factors) for completeness.
 
 \begin{equation}
 \left( {\bf S}_{\nu ^{\prime }=3,4}^{-1}\right) _{r,s}^{r^{\prime
 },s^{\prime }}=-\frac 4{\sqrt{p(p+2)}}\sin \pi \left( rr^{\prime }\alpha
 _{+}^2-\frac{rs^{\prime }}2\right) \sin \pi \left( ss^{\prime }\alpha
_{-}^2-%
 \frac{sr^{\prime }}2\right)  \label{u}
 \end{equation}
 
 \begin{equation}
 \left( {\bf S}_{\nu ^{\prime }=2}^{-1}\right) _{r,s}^{r^{\prime
},s^{\prime
 }}=-\frac{4\gamma _{r^{\prime },s^{\prime }}}{\sqrt{p(p+2)}}\sin \pi
\left(
 rr^{\prime }\alpha _{+}^2-\frac{sr^{\prime }}2\right) \sin \pi \left(
 ss^{\prime }\alpha _{-}^2-\frac{rs^{\prime }}2\right)  \label{uu}
 \end{equation}
 
 \begin{equation}
 \left( {\bf S}_{\nu ^{\prime }=1}^{-1}\right) _{r,s}^{r^{\prime
},s^{\prime
 }}=-\frac 4{\sqrt{p(p+2)}}\sin \pi \left( rr^{\prime }\alpha
_{+}^2-\frac{%
 sr^{\prime }}2+\frac{r^{\prime }}2\right) \cos \pi \left( ss^{\prime
}\alpha
 _{-}^2-\frac{rs^{\prime }}2-\frac{r^{\prime }}2\right)  \label{uuu}
 \end{equation}
 
 with 
 \begin{equation}
 \gamma _{r^{\prime },s^{\prime }}=1-\frac 12\delta _{p,2{\bf {Z}}}\delta
 _{r^{\prime },\frac p2}\delta _{s^{\prime },\frac{p+2}2}  \label{e6}
 \end{equation}
 Recall that the physical origin of the factor $\gamma _{r^{\prime 
},s^{\prime }}$ is
the
 double degeneracy of the R states, apart from the vacuum $(r^{\prime
 },s^{\prime })=({\frac p2},{\frac{{p+2}}2})$.
 
 These equations are essentially the modular transformations of
 superconformal characters \cite{matsuo} except for ${\bf S^{\tilde
 R\rightarrow \tilde R}}$ ~or~ ${\bf S}_{\nu ^{\prime }=1}^{-1}$ which 
is not defined except for the
$(r,s)=\left(
 \frac p2,\frac{p+2}2\right) $ state where it is the identity.
 
 \section{The Verlinde theorem in N=1 superconformal models.}
 
 Even though the proof of the Verlinde theorem \cite{verlinde},\cite
 {verlinde2}, \cite{moore} requires that left and right extended chiral
 algebras consist only of generators with integral conformal weight, it
was
 shown in \cite{aann} that it is possible to construct a complete Verlinde
 basis in N=1 superconformal minimal models, namely 
 \begin{equation}
 {\bf G}=(E_{+}~E_{-}~O_{+}~O_{-})  \label{vb}
 \end{equation}
 with 
 \begin{eqnarray}
 E_{\pm }(r,s) &=&{\frac 12}[G_1^{\nu =3}(r,s)\pm e^{-i\pi ({\frac{\lambda
^2%
 }{2N}}-{\frac 1{12}})}G_1^{\nu =4}(r,s)],  \nonumber \\
 O_{\pm }(r,s) &=&{\frac 12}[G_1^{\nu =2}(r,s)\pm e^{{\frac{i\pi
}4}}G_1^{\nu
 =1}(r,s)]
 \end{eqnarray}
 which is an eigenstate of $\phi (a)$ and with respect to which $\phi (b)$
 yields the fusion rule coefficients. In this basis the descendants in a
$q$
 expansion are always at integer level spacing above the highest weight
state
 (regarding superdescendants as Virasoro primaries). Moreover, the proof of
 the Verlinde theorem relies only on conformal and duality properties
under
 certain manipulations of conformal blocks in their degeneration and
 factorization limits. Since we have checked that these limits are indeed
 consistent with the fusion rules, this suggests that by working in the
 appropriate basis the proof can always be carried through. Even though we
 have no general proof, we will show in this section that the generalized
 Verlinde formula gives the correct fusion coefficients in the TIM and AT
 models, including the fermionic parity of the fields.
 
 Let us summarize the arguments leading to the generalization of the
Verlinde
 formula. The fusion coefficients $N_{IJ}^K$ are defined as 
 \begin{equation}
 {\varphi }_I\times {\varphi }_J=N_{IJ}^K~{\varphi }_K
 \end{equation}
 where ${\varphi }_I$ is one of the operators $N_\alpha ^B$, $N_\alpha ^F$
or 
 $R_\alpha ^{\pm }$, and the upper case indices denote both $r,s$ and spin
 structure sector $\nu $ (or, rather, the appropriate combinations of spin
 structures discussed above). As shown in \cite{aann}, the $b$ cycle
 monodromy operator in the $t$ channel, $\phi ^t(b)$, yields the expected
 superconformal fusion rules. Since we are interested in the action of
$\phi
 _I(b)$ on characters rather than two-point blocks, we may factorize the
$t$
 channel blocks on the identity intermediate state. In general one has an
 equation of the form 
 \begin{equation}
 \phi _I(b)\chi _J=\sum_KN_{IJ}^K\chi _K
 \end{equation}
 where $\chi _I$ denotes a particular character (or combination of
 characters) and the normalization condition is $N_{I0}^K=\delta _I^K$,
where 
 $J=0$ denotes the identity or NS vacuum character. Furthermore, with
respect
 to the same basis of conformal blocks, one has under $\phi ^t(a)$ 
 \begin{equation}
 \phi _I(a)\chi _J=\lambda _I^{(J)}\chi _J
 \end{equation}
 
 One may now proceed as in \cite{verlinde}, using the conjugation relation
 between $a$ and $b$ cycle monodromy, $\phi ^t(a)=S^{-1}\phi ^t(b)S$, to
 express the fusion coefficients in terms of the modular matrices and the
 eigenvalue of $\phi _I(a)$: 
 \begin{equation}
 N_{IJ}^K=\sum_LS_J^L\lambda _I^{(L)}(S^{-1})_L^K
 \end{equation}
 The modular matrix ${\cal S}_I^K$ acts on the Verlinde basis (\ref{vb})
as 
 \begin{equation}
 {\bf G}(r,s|\tau )=\sum_{r^{\prime },s^{\prime }}({\cal
S})_{r,s}^{r^{\prime
 },s^{\prime }}{\bf G}(r^{\prime },s^{\prime }|-{\frac 1\tau })
 \end{equation}
 and is given by 
 \begin{equation}
 {\cal S=}\left( 
 \begin{array}{cccc}
 {\bf S}_{\nu ^{\prime }=3}^{-1} & {\bf S}_{\nu ^{\prime }=3}^{-1} & {\bf
S}%
 _{\nu ^{\prime }=2}^{-1} & {\bf S}_{\nu ^{\prime }=2}^{-1} \\ 
 {\bf S}_{\nu ^{\prime }=3}^{-1} & {\bf S}_{\nu ^{\prime }=3}^{-1} & -{\bf
S}%
 _{\nu ^{\prime }=2}^{-1} & -{\bf S}_{\nu ^{\prime }=2}^{-1} \\ 
 {\bf S}_{\nu ^{\prime }=4}^{-1} & -{\bf S}_{\nu ^{\prime }=4}^{-1} & {\bf
S}%
 _{\nu ^{\prime }=1}^{-1} & -{\bf S}_{\nu ^{\prime }=1}^{-1} \\ 
 {\bf S}_{\nu ^{\prime }=4}^{-1} & -{\bf S}_{\nu ^{\prime }=4}^{-1} &
-{\bf S}%
 _{\nu ^{\prime }=1}^{-1} & {\bf S}_{\nu ^{\prime }=1}^{-1}
 \end{array}
 \right)   \label{sm}
 \end{equation}
 Using the normalization condition allows the eigenvalue to be expressed
as 
 \begin{equation}
 \lambda _I^{(L)}={\frac{S_I^L}{S_{I=0}^L}}
 \end{equation}
 whence 
 \begin{equation}
 N_{IJ}^K=\sum_L{\frac{{S_I^LS_J^L(S^{-1})_L^K}}{S_{I=0,L}}}
 \end{equation}
 Since ${\cal S}^2=1$, multiplying by ${\cal S}^2$ on the right yields the
 result for the number of couplings between three operators labeled by
$I,J,K$%
 : 
 \begin{equation}
 N_{IJK}=\sum_L{\frac{{S_{I,L}S_{J,L}S_{L,K}}}{S_{I=0,L}}}  \label{fc}
 \end{equation}
 Notice that the ${\cal S}$ matrix (\ref{sm}) is symmetric and unitary for
$p$
 odd. This is in accord with the well known result that for $p=3$ in
 particular, there are two equivalent representations of the TIM- either
as a 
 $p=4$ minimal conformal model or as the $N=1$ superconformal model that we are
 discussing. In fact, for $p=3$ the combinations of blocks (\ref{vb}) in
the
 factorization limit are the Virasoro characters 
 \begin{eqnarray}
 E_{+}(1,1) &\sim &\chi _0^{Vir}\qquad E_{-}(1,1)\sim \chi _{\frac
32}^{Vir} 
 \nonumber \\
 E_{+}(1,3) &\sim &\chi _{\frac 1{10}}^{Vir}\qquad E_{-}(1,3)\sim \chi
 _{\frac 35}^{Vir}  \nonumber \\
 O_{\pm }(2,3) &\sim &\chi _{\frac 3{80}}^{Vir}\qquad O_{\pm }(2,1)\sim
\chi
 _{\frac 7{16}}^{Vir}  \label{p}
 \end{eqnarray}
 
 However for $p$ even the factor $\gamma _{r,s}$ in the matrix ${\bf 
S}_{\nu
 ^{\prime }=2}^{-1}$, associated with the Ramond vacuum state, and the
 vanishing of ${\bf S}_{\nu ^{\prime }=1}^{-1}$ for $(r,s)=\left( \frac
p2,%
 \frac{p+2}2\right) $, appear to prevent ${\cal S}$ from being either
 symmetric or unitary, unlike in the conformal case.
 
 Let us compute the fusion coefficients in the first two cases of the
$N=1$
 superconformal minimal series, $p=3$ and 4. Notice that in equation
(\ref{fc}%
 ) $I=0$ corresponds to $E_{+}(1,1)$ and $%
 L=E_{+}(r,s),E_{-}(r,s),O_{+}(r,s),O_{-}(r,s)$. We have to consider the
 three possibilities $NS\times NS\sim NS$, $R\times R\sim NS$ and
$NS\times
 R\sim R$. In the first case we find, for $p=3,$ i.e. the TIM, 
 \begin{eqnarray}
 N_{(1,3,\pm)(1,3,\pm)(1,1,+)} &=&N_{(1,3,\pm)(1,3,\pm)(1,3,-)}=1
\nonumber \\
 N_{(1,3,+)(1,3,-)(1,1,-)} &=&N_{(1,3,+)(1,3,-)(1,3,+)}=1
\nonumber \\
 N_{(1,1,-)(1,1,-)(1,1,+)} &=&N_{(1,3,\pm)(1,1,-)(1,3,\mp)}=1
\end{eqnarray}
 where we have written only the non-trivial nonvanishing 
 coefficients. They correspond respectively to the conformal 
 fusion rules \cite{bpz,francesco}
 \[
 {\frac 1{10}}\times {\frac 1{10}}=0+{\frac 35},\qquad {\frac 35}\times {%
 \frac 35}=0+{\frac 35},\qquad {\frac 1{10}}\times {\frac 35}={\frac
32}+{\frac 1{10}} 
 \]
 \begin{equation}
 {\frac 1{10}}\times {\frac 32}={\frac 35},\qquad {\frac 35}\times {%
 \frac 32}={\frac 1{10}},\qquad {\frac 32}\times {\frac 32}=0
 \end{equation}
 and contain the information about the fermionic parity of the fields in
 accord with previous results \cite{sotkov}.
 
 Let us now compute the fusions $R\times R\sim NS$. The nonvanishing
 coefficients are 
 \begin{eqnarray}
 N_{(2,3,\pm )(2,3,\pm )(1,1,+)} &=&N_{(2,3,\pm )(2,3,\pm )(1,3,-)}=1 
\nonumber \\
 N_{(2,3,\pm )(2,3,\mp )(1,1,-)} &=&N_{(2,3,\pm )(2,3,\mp )(1,3,+)}=1
\nonumber \\
 N_{(2,1,\pm)(2,3,\pm)(1,3,-)} &=&N_{(2,1,\pm)(2,3,\mp)(1,3,+)}=1
\nonumber \\
 N_{(2,1,\pm)(2,1,\pm)(1,1,+)} &=&N_{(2,1,\pm)(2,1,\mp)(1,1,-)}=1
 \label{aa}
 \end{eqnarray}
These should be compared to the conformal fusion rules \cite{bpz,francesco}  
 \begin{eqnarray}
 {\frac 3{80}}\times {\frac 3{80}}&=&0+{\frac 32}+{\frac 1{10}}+{\frac 35}
\nonumber \\
 {\frac 7{16}}\times {\frac 7{16}}=0+{\frac 32}&,& \quad {\frac 7{16}}\times
{\frac 3{80}}={\frac 1{10}}+{\frac 35}
 \label{aaa}
 \end{eqnarray}

 From the first two lines in eq. (\ref{aa}) we get
 
 \begin{equation}
 \lbrack (2,3,+)+(2,3,-)]\times [(2,3,+)+(2,3,-)]=2\left(
 (1,1,+)+(1,1,-)+(1,3,+)+{(1,3,-)}\right) 
 \end{equation}
 and the first equality in eq. (\ref{aaa}) is reproduced by making the
identification
 
 \begin{equation}
 \frac 1{\sqrt{2}}[(2,3,+)+ (2,3,-)]=\left( \frac 3{80}\right) 
 \end{equation}
 where a $\frac 1{\sqrt{2}}$ factor has been introduced since the double
 degeneracy of the $R$ states requires the identification of two $R$
states
 with only one state in the conformal case (see eq. (\ref{p})).
The analogous identification for $O_{\pm}(2,1)$ allows to reproduce the 
remaining fusions (\ref{aaa}).

 Similarly we obtain for the mixed fusion rules $R\times NS\sim R$ 
 \begin{eqnarray}
 N_{(2,1,\pm)(1,3,\pm)(2,3,-)} &=&N_{(2,1,\pm)(1,3,\mp)(2,3,+)}=1
\nonumber \\
 N_{(2,3,\pm)(1,3,\pm)(2,3,-)} &=&N_{(2,3,\pm)(1,3,\mp)(2,3,+)}=1
\nonumber \\
 N_{(2,3,\pm)(1,3,\pm)(2,1,+)} &=&N_{(2,3,\pm)(1,3,\mp)(2,1,-)}=1
\nonumber \\
 N_{(2,3,\pm)(1,1,-)(2,3,\mp)} &=&N_{2,1,\pm)(1,1,-)(2,1,\mp)}=1
 \label{cosaa}
 \end{eqnarray}
which correspond to \cite{bpz,francesco}
 \[
 {\frac 7{16}}\times {\frac 1{10}}={\frac 7{16}}\times {\frac 35}={%
 \frac 3{80}}
 \]
 \[
 {\frac 3{80}}\times {\frac 1{10}}={\frac 3{80}}\times {\frac 35}={%
 \frac 3{80}}+{\frac 7{16}}
 \]
 \begin{eqnarray}
 {\frac 3{80}}\times {\frac 32}={\frac 3{80}}&,& \quad {%
 \frac 7{16}}\times {\frac 32}={\frac 7{16}}
 \label{cosab}
 \end{eqnarray}
 For example, the first line in eq. (\ref{cosaa}) corresponds to 
 \begin{eqnarray}
 \lbrack (2,1,+)+(2,1,-)]\times \frac 1{10} &=&[(2,3,+)+(2,3,-)] 
\nonumber \\
 \lbrack (2,1,+)+(2,1,-)]\times \frac 35 &=&[(2,3,+)+(2,3,-)]
 \end{eqnarray}
in accord with the first line of eq. (\ref{cosab}), and analogously it is
possible to reproduce the remaining fusions.

 In the $p=4$ case the formalism is completely analogous. The
identification
 among the superconformal \cite{yang2} and the Virasoro \cite{verlinde3}
 characters is as follows
 
 \begin{eqnarray}
 \chi _0^{NS} &=&\chi _0^{Vir}+\chi _{\frac 32}^{Vir,(1)}\qquad \chi _0^{%
 \stackrel{\sim }{NS}}=e^{-i\frac \pi {12}}\left( \chi _0^{Vir}-\chi
_{\frac
 32}^{Vir,(1)}\right)  \nonumber \\
 \chi _1^{NS} &=&\chi _1^{Vir}+\chi _{\frac 32}^{Vir,(2)}\qquad \chi _1^{%
 \stackrel{\sim }{NS}}=e^{-i\frac \pi {12}}\left( \chi _1^{Vir}-\chi
_{\frac
 32}^{Vir,(2)}\right)  \nonumber \\
 \chi _{\frac 16}^{NS} &=&\chi _{\frac 16}^{Vir}+\chi _{\frac
23}^{Vir}\qquad
 \chi _{\frac 16}^{\stackrel{\sim }{NS}}=e^{i\frac \pi 4}\left( \chi
_{\frac
 16}^{Vir}+\chi _{\frac 23}^{Vir}\right)  \nonumber \\
 \chi _{\frac 1{16}}^{NS} &=&\chi _{\frac 1{16}}^{Vir,(1)}+\chi _{\frac
 9{16}}^{Vir,(1)}\qquad \chi _{\frac 1{16}}^{\stackrel{\sim
}{NS}}=e^{i\frac
 \pi {12}}\left( \chi _{\frac 1{16}}^{Vir,(1)}-\chi _{\frac
 9{16}}^{Vir,(1)}\right)  \nonumber \\
 \chi _{\frac 1{24}}^R &=&\chi _{\frac 1{24}}^{Vir}+\chi _{\frac{25}{24}%
 }^{Vir}\qquad \chi _{\frac 1{24}}^{\tilde R}=e^{i\alpha }\left( \chi
_{\frac
 1{24}}^{Vir}-\chi _{\frac{25}{24}}^{Vir}\right)  \nonumber \\
 \chi _{\frac 38}^R &=&\chi _{\frac 38}^{Vir}\qquad \chi _{\frac
 1{16}}^R=\chi _{\frac 1{16}}^{Vir,(2)}\qquad \chi _{\frac 9{16}}^R=\chi
 _{\frac 9{16}}^{Vir,(2)}
 \end{eqnarray}
where $\alpha \in {\bf R}$, the $\chi _{\frac 1{24}}^{\tilde R}$
character
 is an arbitrary phase, and we have taken into account that in the
RCFT description as a $Z_2$ orbifold of a Gaussian model with rational radius,
there are two representations of $\left[ \frac 32\right] ,$ $\left[
 \frac 1{16}\right] $ and $\left[ \frac 9{16}\right] .$
 
 As in the $p=3$ situation, all the doubly degenerate states of the
$R$
 sector must include a $\frac 1{\sqrt{2}}$ factor to be compared with the
 corresponding states in the conformal case. For the $\frac 1{24}$ state
we
 must take into account that in the $\nu =1$ sector our correlation
functions
 cannot reproduce the corresponding character in the factorization limit
 because the intermediate state is a fermion instead of the identity. So
we
 have to introduce a $\frac 1{\sqrt{2}}$ factor here again to take into
 account the contribution of the state $\frac 1{24}$ in both the $\nu =1$
and 
 $\nu =2$ sectors. This corresponds to the states $\frac 1{24}$ and
$\frac{25%
 }{24}$ in the Gaussian model case. Following the same procedure as in the
$p=3$
 case it is possible to see that the fusion rules of the conformal case,
as
 calculated in \cite{verlinde3}, are reproduced. Let's consider, as an
 example, the $\frac 1{24}\times \frac 1{24}$ fusion rule. We obtain
 
 \begin{eqnarray}
 \lbrack (2,3,+)+(2,3,-)]\times [(2,3,+)+(2,3,-)] = \nonumber\\
 4\left((1,1,+)+(1,1,-)
 +(3,1,+)+(3,1,-)+(1,3,+)+(1,3,-)\right) 
 \end{eqnarray}
 in complete agreement, after rescaling properly, with the addition of the
 conformal fusion rules \cite{verlinde3}
 
 \begin{equation}
 \frac 1{24}\times \frac 1{24}=\frac{25}{24}\times \frac{25}{24}=0+1+\frac
 16\quad \quad \frac 1{24}\times \frac{25}{24}=\frac{25}{24}\times \frac
 1{24}=\frac 23+2\left( \frac 32\right)
 \end{equation}
 
 Notice that these fusion rules disagree with those in App. E of ref. \cite
 {aann} (where the set of eqs. (5.9) of that work were used).
 
In conclusion we have provided strong evidence
 that by working in an appropriate basis
 the Verlinde theorem carries through to the N=1 superconformal minimal models.
We have also shown that the generalized Verlinde formula contains information
about the fermionic parity of the fields.

 \end{document}